# Modeling and analysis of the dielectric properties of composite materials using large random RC networks

M.Aouaichia[1,*] and R.Bouamrane[1]

*1. LEPM, Faculté de Physique, USTO-MB, BP 1505 El M'Naouer, Oran 31000, Algeria*

*corresponding author e-mail: mus.aouaichia@gmail.com

**Abstract**

The paper proposes a simple and efficient method to study the dielectric properties of composite materials modeled using very large random resistor-capacitor (RC) networks. The algorithm is based on the Frank-Lobb reduction scheme and analyses the frequency dependent AC conductivity, permittivity and phase angle of composite materials. The simulation study is based on 1056 samples of random network containing $2 \times 10^6$ components randomly positioned in different proportion of capacitors. It has been found that these properties exhibit similarities to the universal dielectric response of dielectric materials. The results show that there is no variability between samples at low and high frequencies and across the power law.

1. **Introduction**

Many materials used in daily life are composite and are often a blend of at least two constituents or phases [1] with conductive and dielectric (insulating) properties. These systems display a similar frequency dependent conductivity and permittivity [2, 3].
The frequency dependent complex conductivity $\tilde{\sigma}$ and permittivity $\tilde{\varepsilon}$ of composite materials are given respectively as [1, 8]:

$$\tilde{\sigma}(\omega) = \sigma(0) + A(i\omega)^n \tag{1}$$

$$\tilde{\varepsilon}(\omega) = B\omega^{n-1} + \varepsilon_\infty \tag{2}$$

Where $\sigma(0)$ is the dc conductivity; $A$ and $B$ are constant; $0 < n > 1.0$; $\varepsilon_\infty$ is the dielectric permittivity which exhibits a frequency independence at high frequencies, $\omega$ is an angular frequency.

These electrical properties obey the anomalous power law [4] well known as the Universal Dielectric Response (UDR) [5, 6, 7]. Examples of universality of behavior of ceramic materials can be found in [2, 3]. Such features are modeled by simulating the electrical properties using a random network of resistors and capacitors [8, 9, 10]. The electrical network represents a heterogeneous microstructure that contains two constituents insulating and conductive regions [2]. An example of this mixture that can be shown as a resistor-capacitor network in Fig. 1(a).

There are a large number of papers in the literature on the application of random RC network in modeling; however in most of those studies the authors use a small number of components (N) of networks and realizations. For example Almond et al. [14] explored a size of around $10^4$ components, 100 realizations. Bouamrane et al. [4] used a network of $32 \times 10^4$ components, 256 realizations. Tuncer et al. [11] studied the response from several circuit cases with $26 \times 10^4$ lattice size and Hamou et al. [12] generated a network of $5 \times 10^5$ components, 10,000 realizations. An example of these results is shown in Fig.2. The random network simulated consists of 512 components with 60%, 40% proportions of resistances and capacitances respectively.

In this work, the authors have developed an algorithm to explore a large random RC network which includes $2 \times 10^6$ components and 1056 realizations. It has been found that for this size all the electrical properties of mixture materials are verified. The main goal of this work is to examine how well the response of conductivity, permittivity and phase angle of random mixtures materials may be modeled by large random network RC.

## 2. The RC Network Model and the Calculation Process
### 2.1 Principle of the numerical calculation

A square network with $N = 2 \times n \times n$ components is used to demonstrate the calculation methodology, where $n$ is the number of nodes. For example, Fig. 1(b) shows a network with n = 5 and thus it has 50 components randomly positioned. The dielectric response $\varepsilon(\omega)$ of a mixture random network R-C is determined by calculating the equivalent impedance or (admittance) from the following relation:

$$\varepsilon(\omega) = \varepsilon'(\omega) + i\varepsilon''(\omega) = 1/i\varepsilon_0 \omega Z_{eq}(\omega) = Y_{eq}(\omega)/i\varepsilon_0 \omega \qquad (3)$$

The equivalent complex impedance $Z_{eq}(\omega)$ is calculated using an efficient algorithm, based on the Frank-Lobb (FL) reduction scheme. The FL method uses a mathematical technique that simplifies the study of some electrical networks. This method consists of applying successive triangle-star Fig. 3(a-b) and star-triangle Fig. 3(c-d) transformations.

The FL reduction scheme starts with the triangle of Fig. 3(a) which is transformed into a star to produce the configuration of Fig. 3(b) following the relation: (Δ → Y)

$$\begin{cases} Y_6 = \frac{Y_1 Y_2 + Y_1 Y_3 + Y_2 Y_3}{Y_3} \\ Y_7 = \frac{Y_1 Y_2 + Y_1 Y_3 + Y_2 Y_3}{Y_2} \\ Y_8 = \frac{Y_1 Y_2 + Y_1 Y_3 + Y_2 Y_3}{Y_1} \end{cases} \qquad (4)$$

Then the star of Fig. 3(c) is transformed back into a triangle as shown in Fig. 3(d) following the relation: (Y → Δ)

$$\begin{cases} Y_9 = \frac{Y_4 Y_8}{Y_4 + Y_5 + Y_8} \\ Y_{10} = \frac{Y_5 Y_8}{Y_4 + Y_5 + Y_8} \\ Y_{11} = \frac{Y_4 Y_5}{Y_4 + Y_5 + Y_8} \end{cases} \qquad (5)$$

The final result of this sequence of transformations is to reduce any large random network (RC) to a single link that have the same equivalent impedance as the entire network.

**2.2 Random network RC characteristics**

The square network (R-C) of Fig. 1(b) contains a random mixture of resistors and capacitors with a proportion of $(1-p)$ resistors ($1\,k\Omega$) and $p$ capacitors ($1\,nF$), with the admittance randomly as either $y_R = 1/R$ or $y_c = i\omega C$ respectively where $\omega$ the angular frequency is $2\pi f$. The critical percolation for a 2D square network is $p_c = 1/2$ [14] and the fundamental study of random RC networks [3, 8, 9, 14] were:

- When the proportion of capacitors is different from the critical proportion $p \neq p_c$:
  i. If $p < p_c$, the overall admittance is independent of $\omega$ ($|Y| \propto y_R$) at low and high frequency, the network is resistive.
  ii. If $p > p_c$, the overall admittance is dependent on $\omega$ ($|Y| \propto y_C$) at low and high frequency.
- When the proportion of capacitors equal critical proportion ($p = p_c$):

i. At low frequency, half of the realizations of the network give an admittance response independent of $\omega$ with a resistive percolation, and half gives an admittance response proportional to $\omega$ with a capacitive percolation. A similar behavior is observed at high frequencies.

- For all cases above, at intermediate values of $\omega$ the admittances of capacitors and resistors become comparable ($y_R \sim y_C$), the emergent power-law behavior is observed and this is characterized by the admittance response $|Y| \propto \omega^p$.
- For the resulting network associated with a simple logarithmic mixing rule, the complex conductivity and permittivity of the network could be described as:

$$\sigma^* = (i\omega C)^p (R^{-1})^{1-p} \propto (i\omega)^p \tag{6}$$

$$\varepsilon^* = \sigma^*/i\omega \propto (i\omega)^{p-1} \propto (i\omega)^{p-1} \tag{7}$$

- The logarithmic mixing rule also predicts that the phase angle in radians, is

$$\phi(\omega) = (1-p).\pi/2 \tag{8}$$

Where $p$ is the fraction of capacitors.

### 3. Result and discussion

To investigate the results obtained in previous work that uses small network sizes and realizations, the network responses of 1056 separate networks of randomly positioned resistors and capacitors were developed, which contains $N = 2097152$ components.

Fig. 4 shows the real $\varepsilon'(\omega)$ and imaginary $\varepsilon''(\omega)$ of permittivity as a function of the angular frequency $\omega$, expressed in log-scale, for $p = p_c = 1/2$. Similar results are shown in Fig. 5 for $p = 0.4 < p_c$, $p = 0.6 > p_c$. The logarithmic mixing rule of Eq. (6-7) used here leads to the power law frequency dependence on the complex conductivity and permittivity [8, 9]. These figures show an excellent dependence with $\omega^p$ and $\omega^{-\vartheta}$, where $p$ and $\vartheta$ is the capacitor and resistor concentration respectively, as explained by Almond and al [9]. These results agree with those of the previous work.

In Fig. 6(a), where the proportion is $R = 60\%, C = 40\%$, the phase of the conductive network at low and high frequencies is $\phi(\omega) \approx 0°$, corresponding to the high and low frequency of admittance Fig. 7(a). At intermediate frequency ($10^4 \sim 10^8$) the power law of both $\varepsilon'$ and $\varepsilon'' \propto \omega^{-0.6}$ is preserved and Eq. (8) gives a phase of $-36°$. The network simulated in Fig. 5 (b) exhibits no peak in dielectric loss $\varepsilon''$ because there is a conductivity of the conductive network.

In Fig. 6(b), where the proportion is $R = 40\%, C = 60\%$, the phase is $\phi(\omega) \approx -90°$ at high and low frequency, corresponding to the high and low frequency of $\varepsilon'$ in Fig. 5(c) where the network becomes purely capacitive. At intermediate frequencies, the phase increases and is approximately constant in the frequency range $(10^4 \sim 10^8)$. In the same frequency range $\varepsilon'$ and $\varepsilon''$ lowering with frequency in power law $\varepsilon'$ and $\varepsilon''$ are proportional to $\omega^{-0.4}$. In this case Eq. (8) gives a phase $\phi(\omega) = -54°$. For the case of R=50%, C=50%, as shown in Fig. 6(c), at low frequency the phase of the network is approximately $-90°$ therefore it is influenced by the high impedances of the capacitors. At the high frequencies the phase approximately $0°$, and the network is influenced by the high impedance of the resistors. At intermediate frequencies the phase equal $\sim 45°$ Fig. 6(c), predicted by the Eq. (8) and both of $\varepsilon'$ and $\varepsilon''$ go with power law proportional to $\omega^{-0.5}$ Fig. 4.

The results of the admittance in cases where the proportions of capacitors $p = 0.4$, $p = 0.6, p = 0.5$ are shown in Fig.7 for 1056 different realizations of large network size of $N = 2097152$ components as a function of $\omega$. By plotting the numerical calculations with the logarithmic mixing rule of Eq. (6), it can be seen that the predictions of this rule perfectly agree with results of the numerical simulations. In all cases, the intermediate frequency power law region is clearly defined with $\omega^p$.

## 4. Conclusions

This work has analyzed very large number of random network resistors-capacitors and the frequency responses of the permittivity, admittance, and phase angle are simulated. The calculations have been performed using an efficient algorithm developed based on the Frank-Lobb reduction scheme. The results of this large network confirm that the power law characteristic is an emergent property of random RC network and the frequency response of permittivity, admittance, and phase angle of composite materials are verified. It has been shown that the permittivity decreases with $\omega^{-\vartheta}$, $\vartheta$ is the proportion of resistance and the admittance increase with $\omega^p$ where $p$ is the proportion id capacitance.  It has been found that frequency ranges through which these power law regions are produced do not increase with the size of the network. The prediction of the logarithmic mixing rule is clearly demonstrated for large random networks RC. For these reasons it can be concluded that the proposed method is valid for networks of any sizes.

**Acknowledgments**

M.Aouaichia and R. Bouamrane would like to thank the Directorate General for Scientific Research and Technology Development (DG-RSDT) for providing our university with a high performance computing facility.

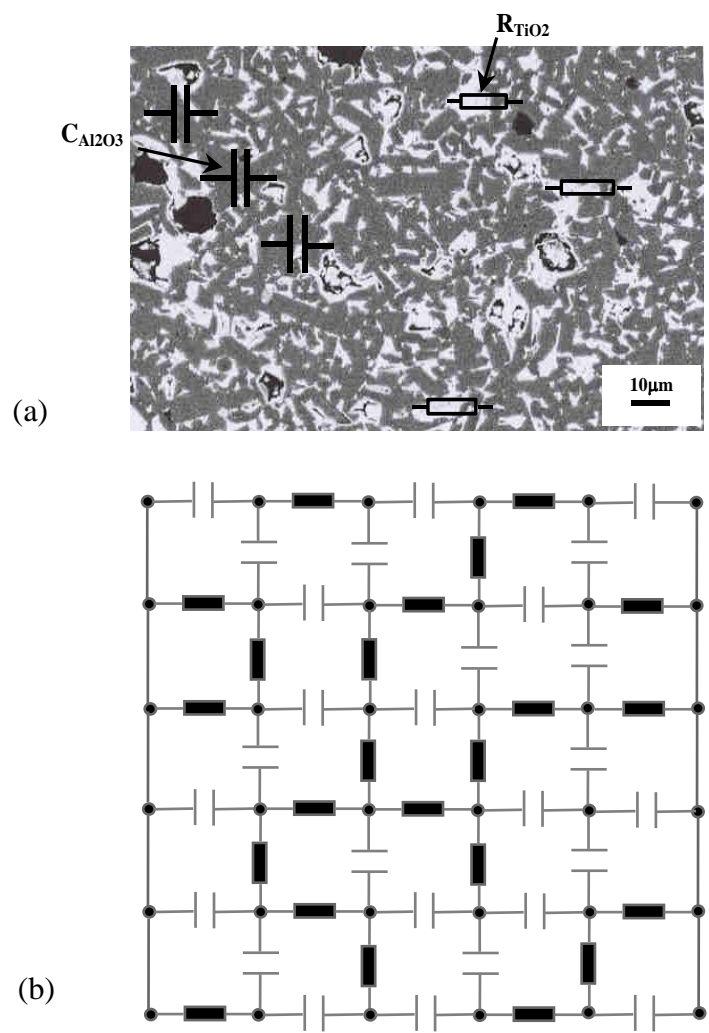

**Fig. 1.** (a) Microstructure of $Al_2O_3 - TiO_2$ composite: grey phase corresponds to ($C_{AL2O3}$, capacitor); white phase corresponds to ($TiO_2$, conductor) [2]. (b) Example of resistor (white phase) capacitors (grey phase) network.

**Figure**

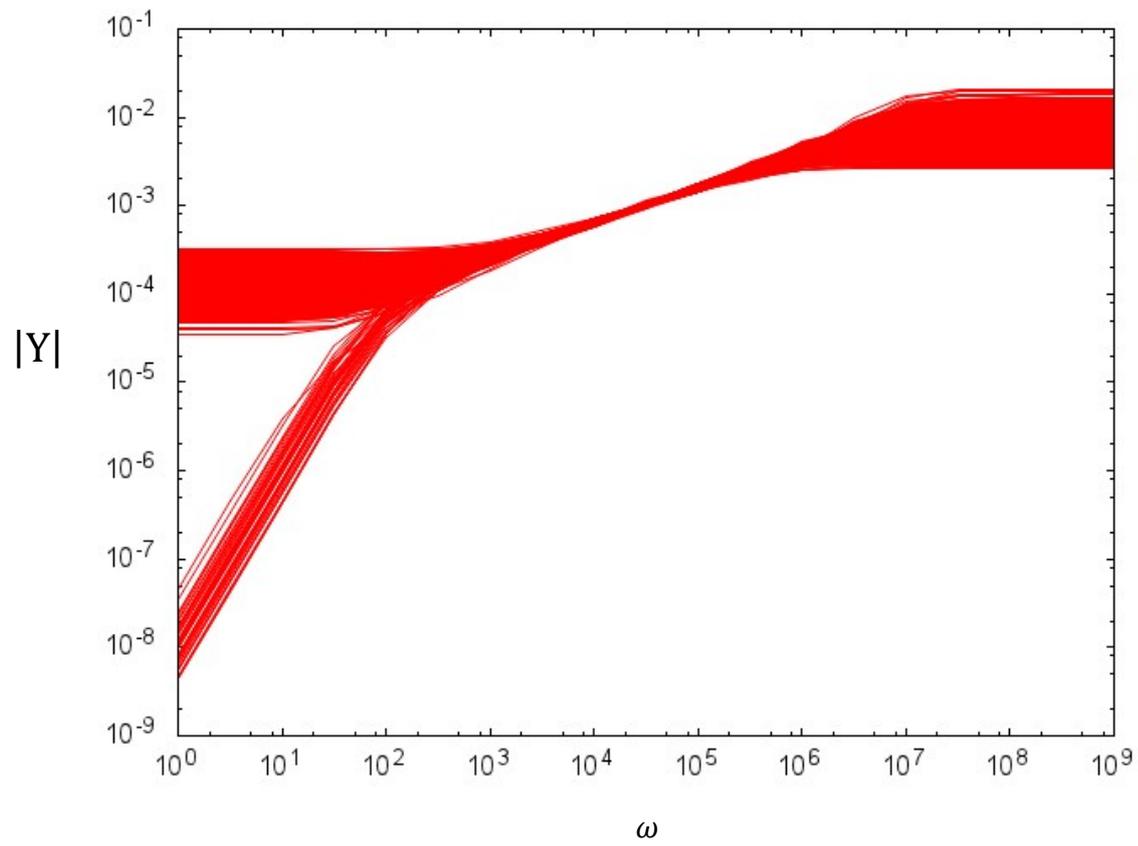

**Fig. 2.** Example of Ac conductivities of 256 R–C networks of 512 components (60% R, 40% C).

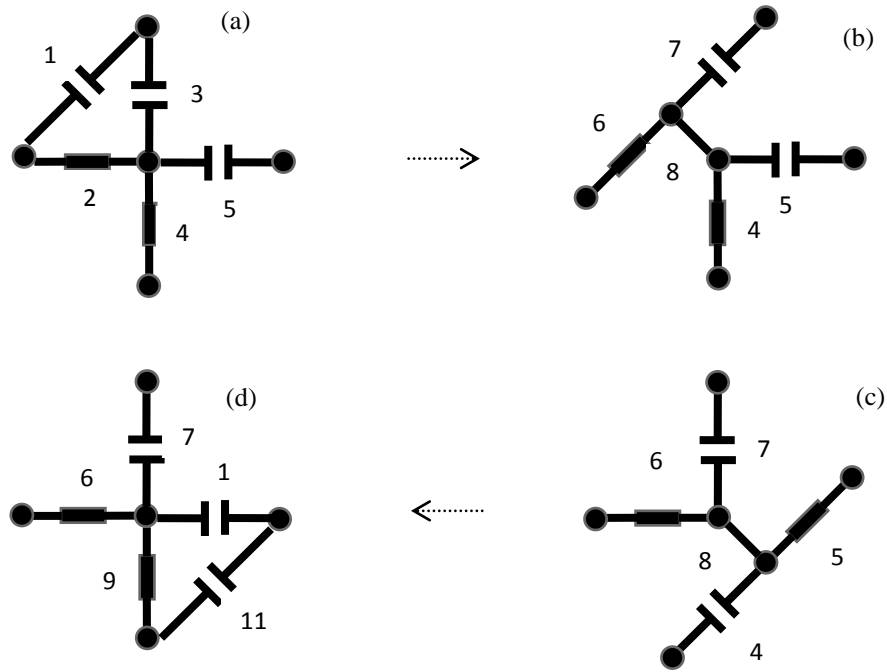

**Fig.3.** Fundamentals transformation: $(a - b) \rightarrow (\Delta - Y), (c - d) \rightarrow (Y - \Delta)$.

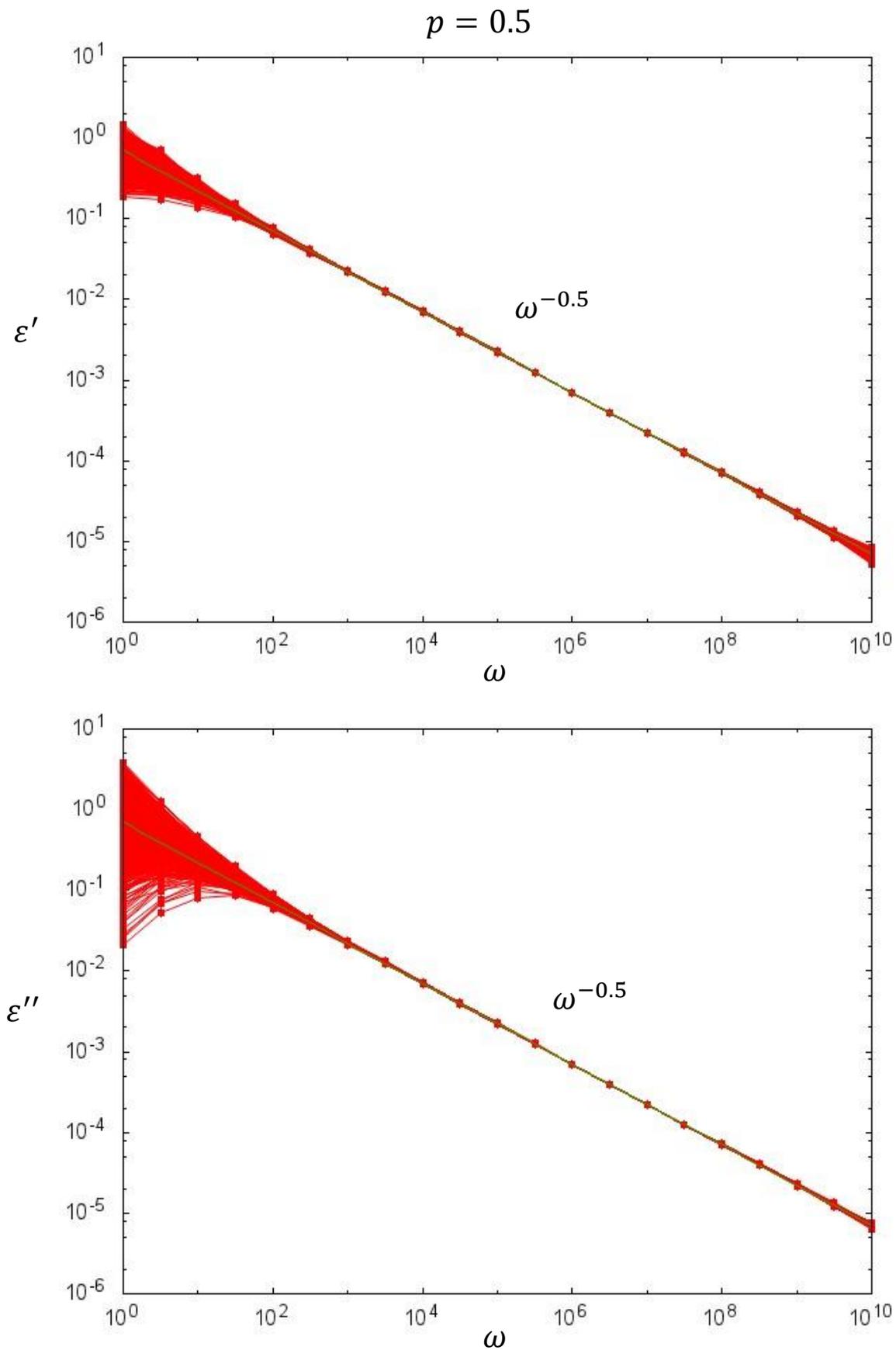

**Fig.4.** Real and imaginary permittivity $\varepsilon'(\omega), \varepsilon''(\omega)$ as a function of $\omega$ with rresponses of network simulation of $p = 1/2$. The results of Fig. 4 are obtained from the simulation of a network containing 2097152 components for 1056 random realizations. The green line shows the logarithmic mixing formula, Eqn (7), prediction for the networks.

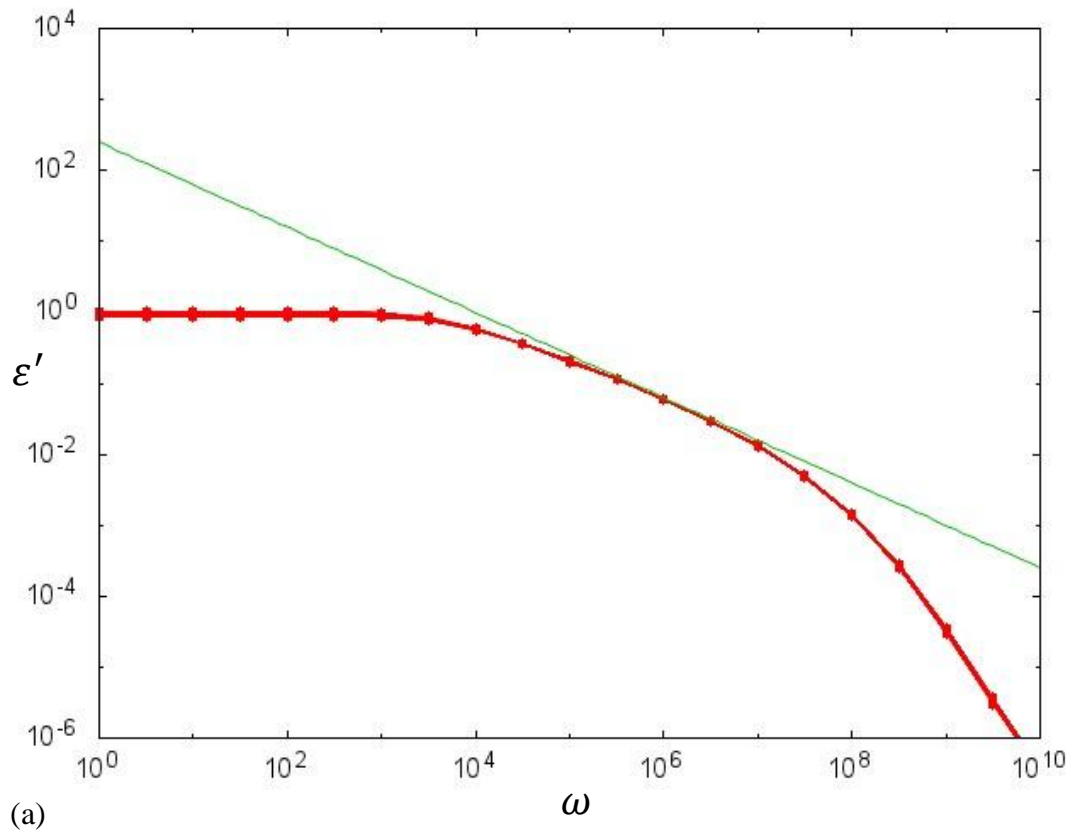

(a)

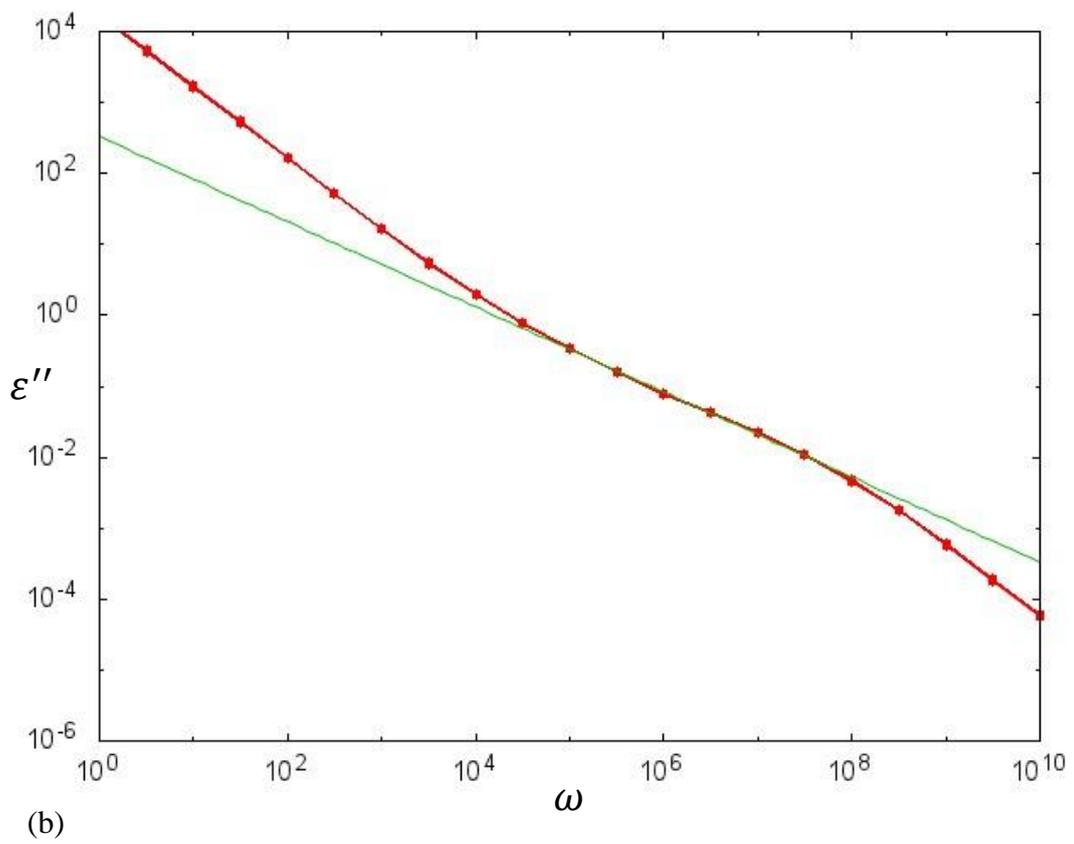

(b)

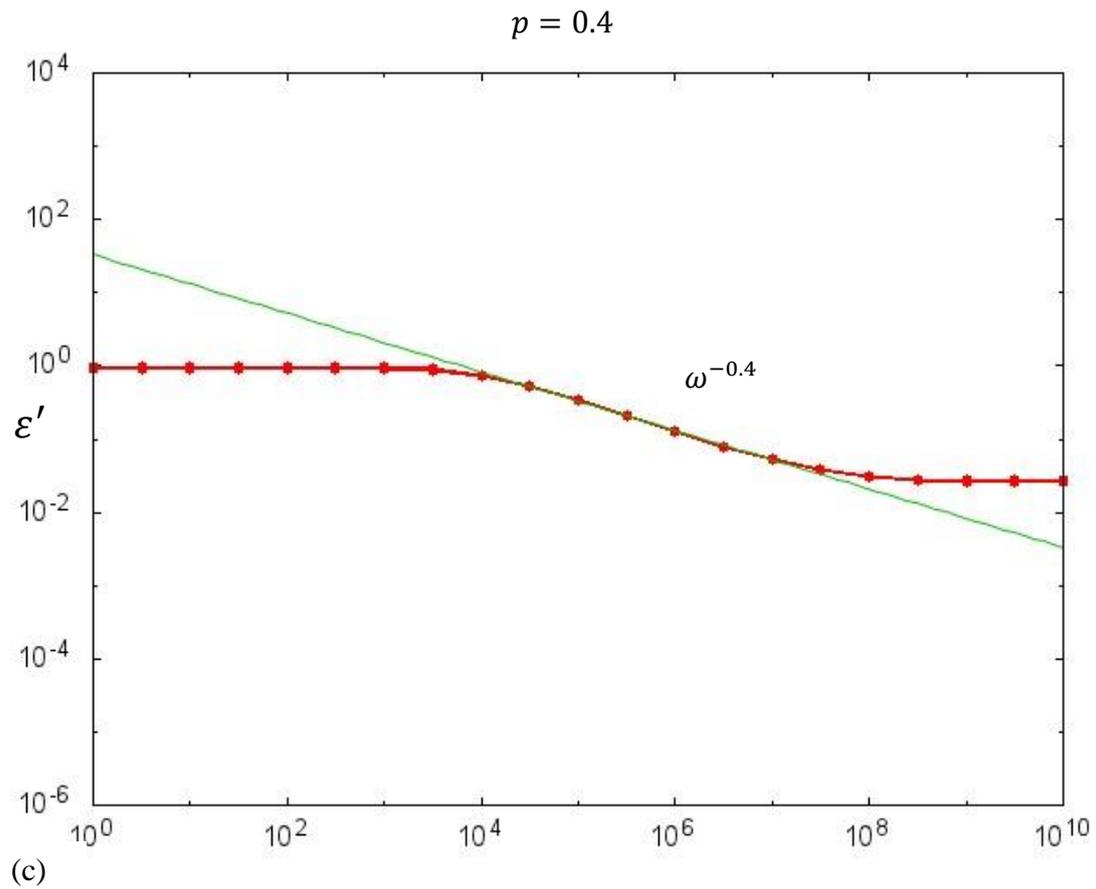

(c)

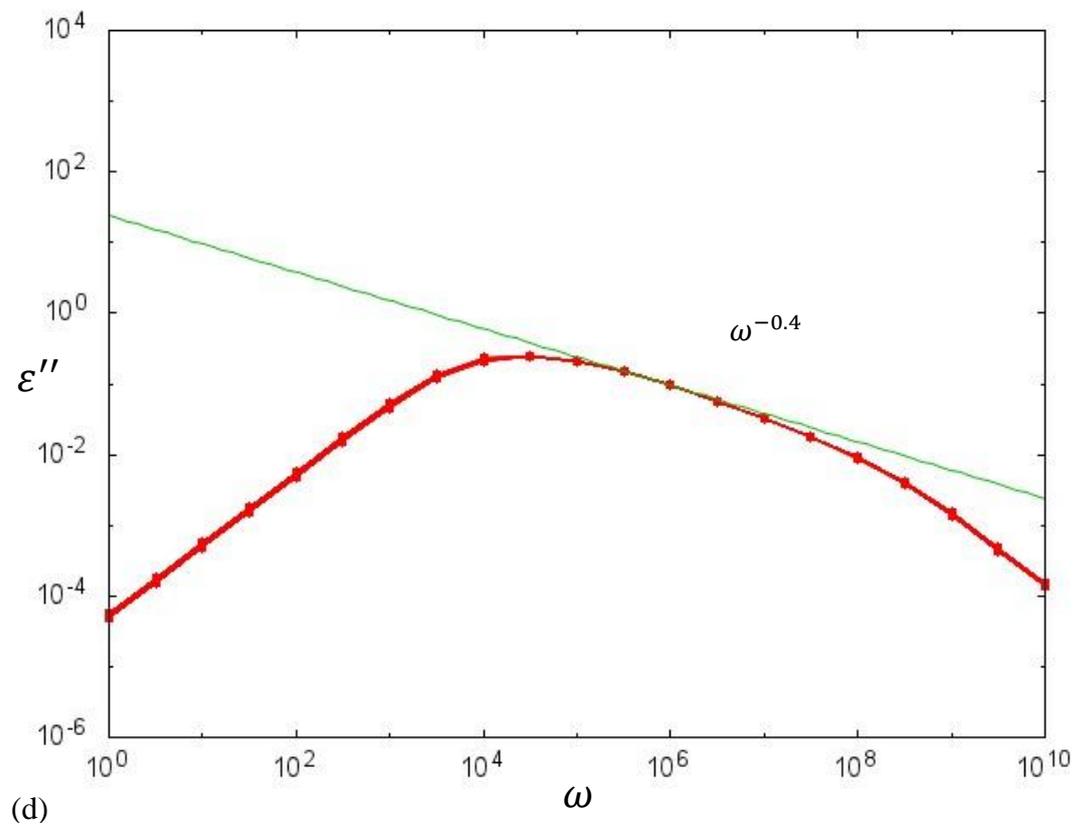

(d)

**Fig. 5.** The real and imaginary permittivity $\varepsilon'(\omega), \varepsilon''(\omega)$ as a function of $\omega$ with Responses of network simulation of $p = 0.6, 0.4$. The results of Fig. 5 are obtained from the simulation of a network containing 2097152 components for 1056 random realizations. The green line shows the logarithmic mixing formula, Equation (7), prediction for the networks.

**Figure**

$p = 0.4$

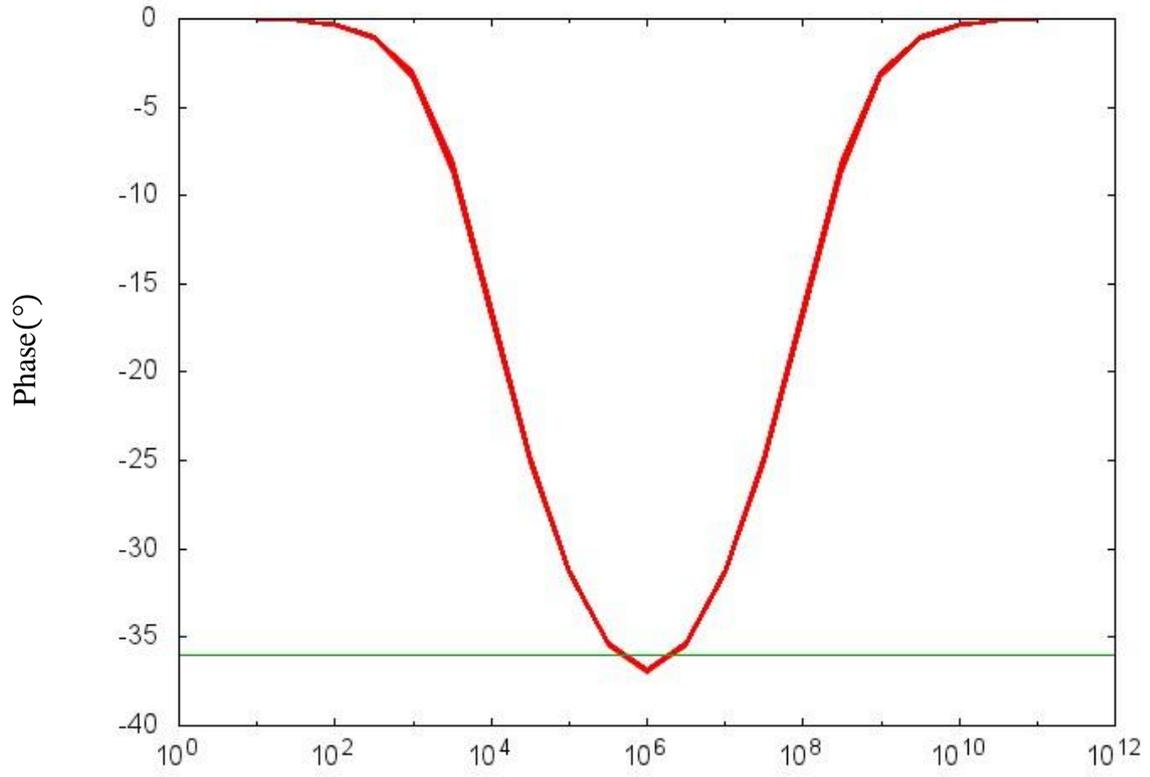

(a)

$p = 0.6$

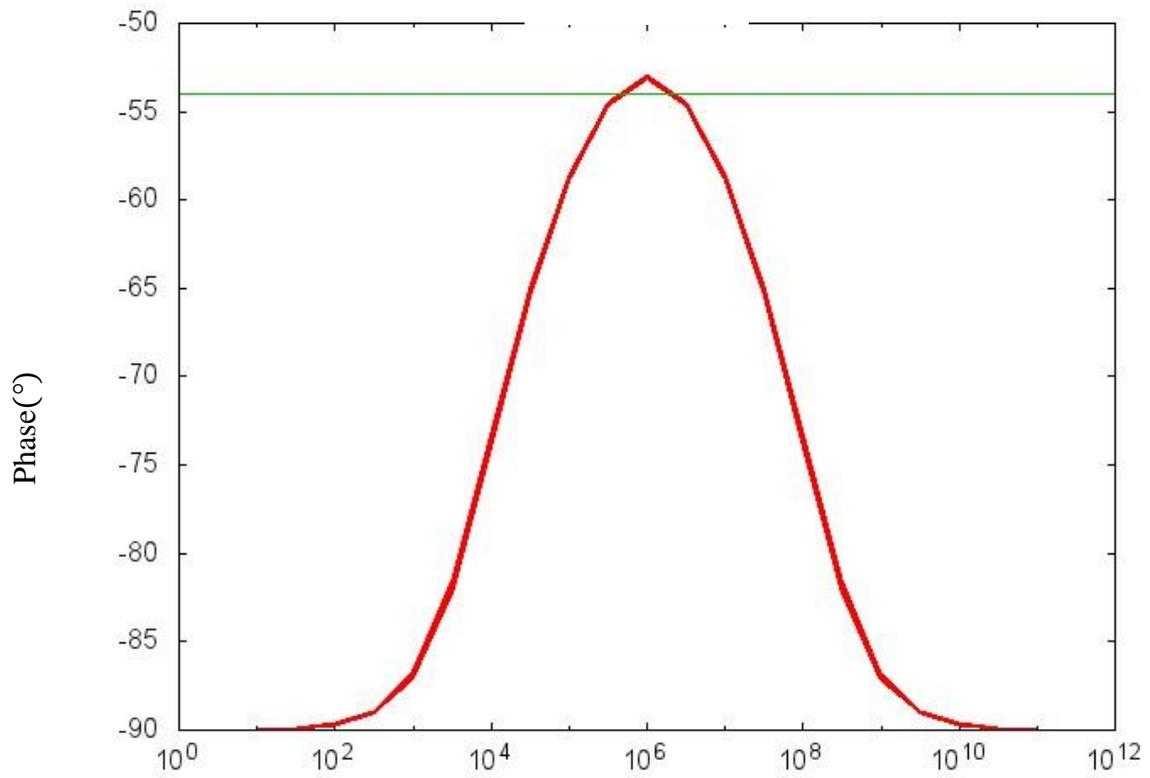

(b)

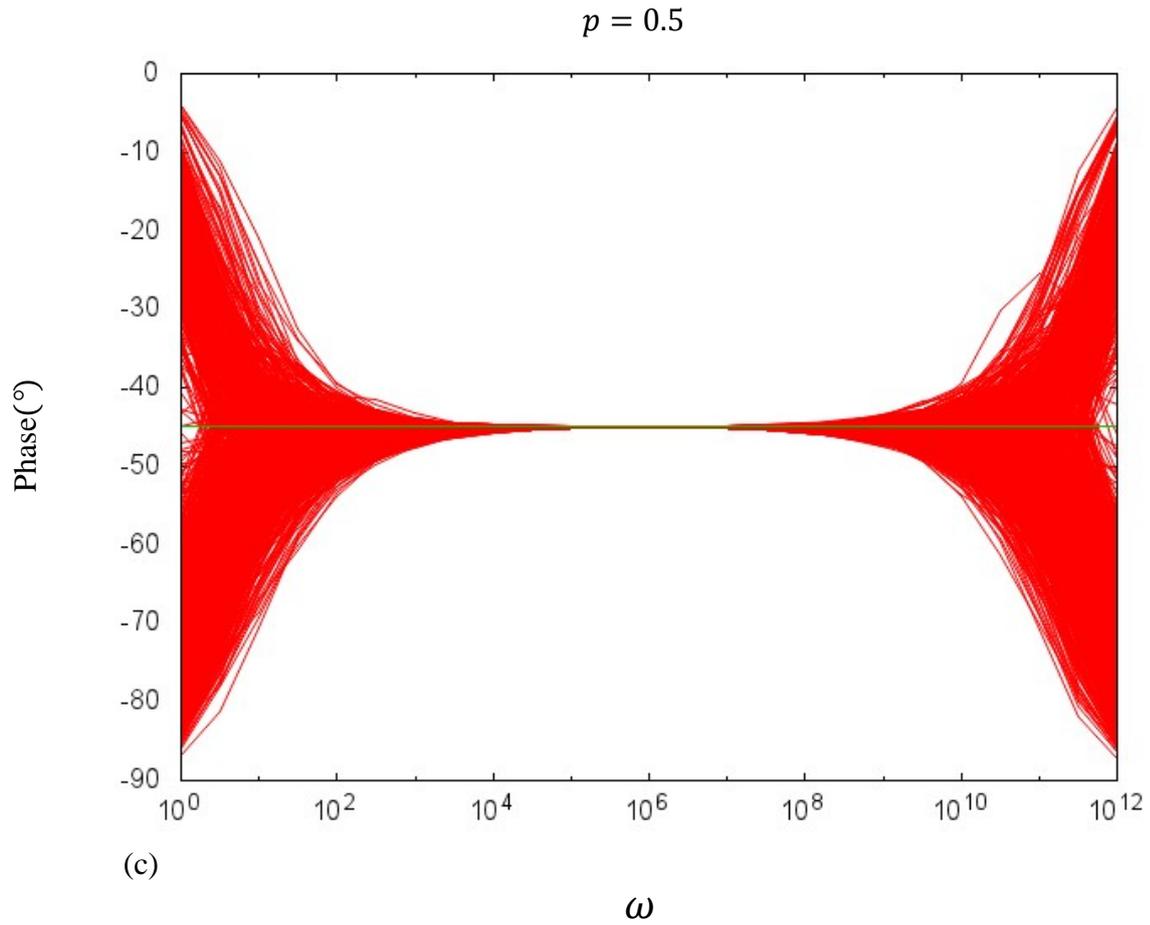

**Fig. 6.** The phase-frequency of network which contains 2097152 components, with response of 1056 realizations for different cases of $p = 0.4, 0.6, 0.5$ respectively. The green line shows the logarithmic mixing formula, Eqn (8), prediction for the networks.

**Figure**

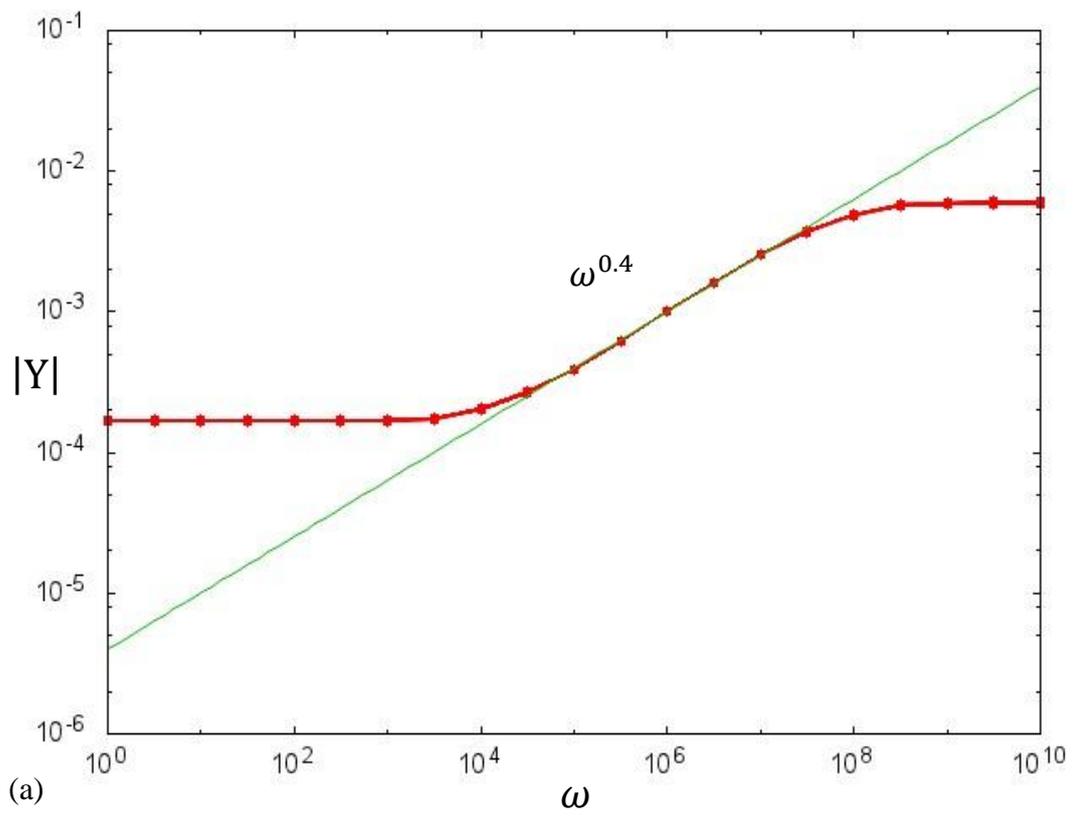

(a)

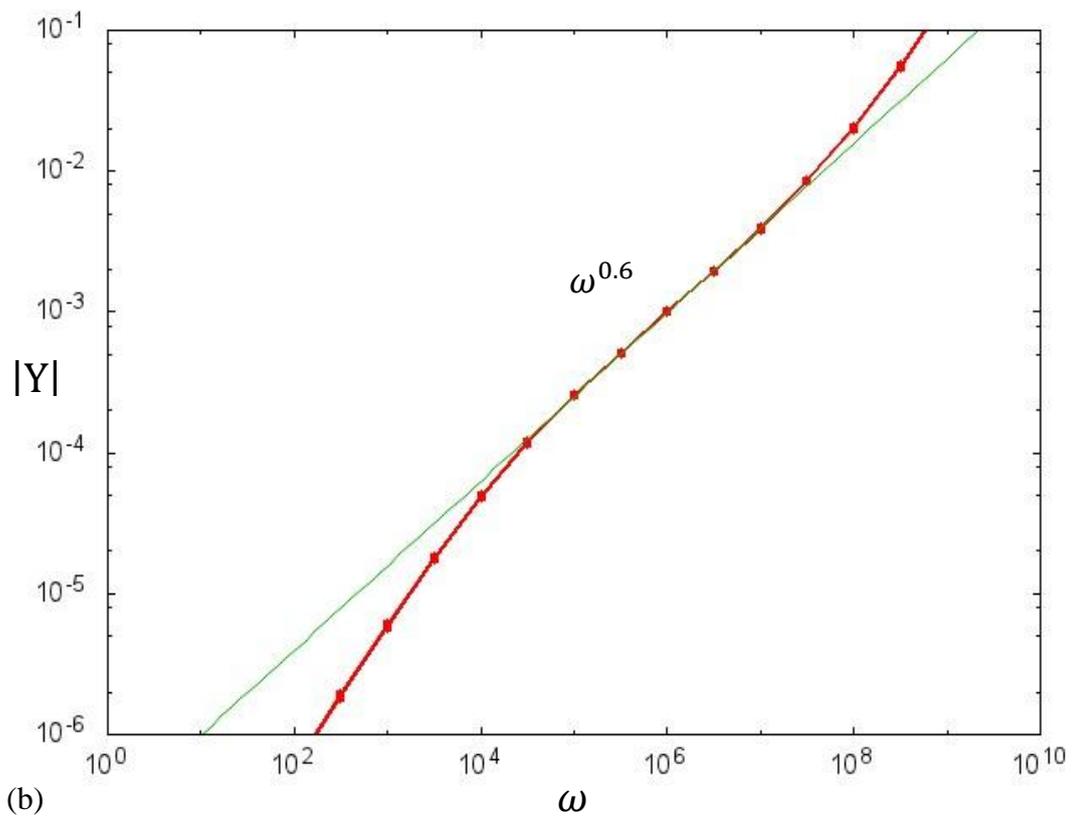

(b)

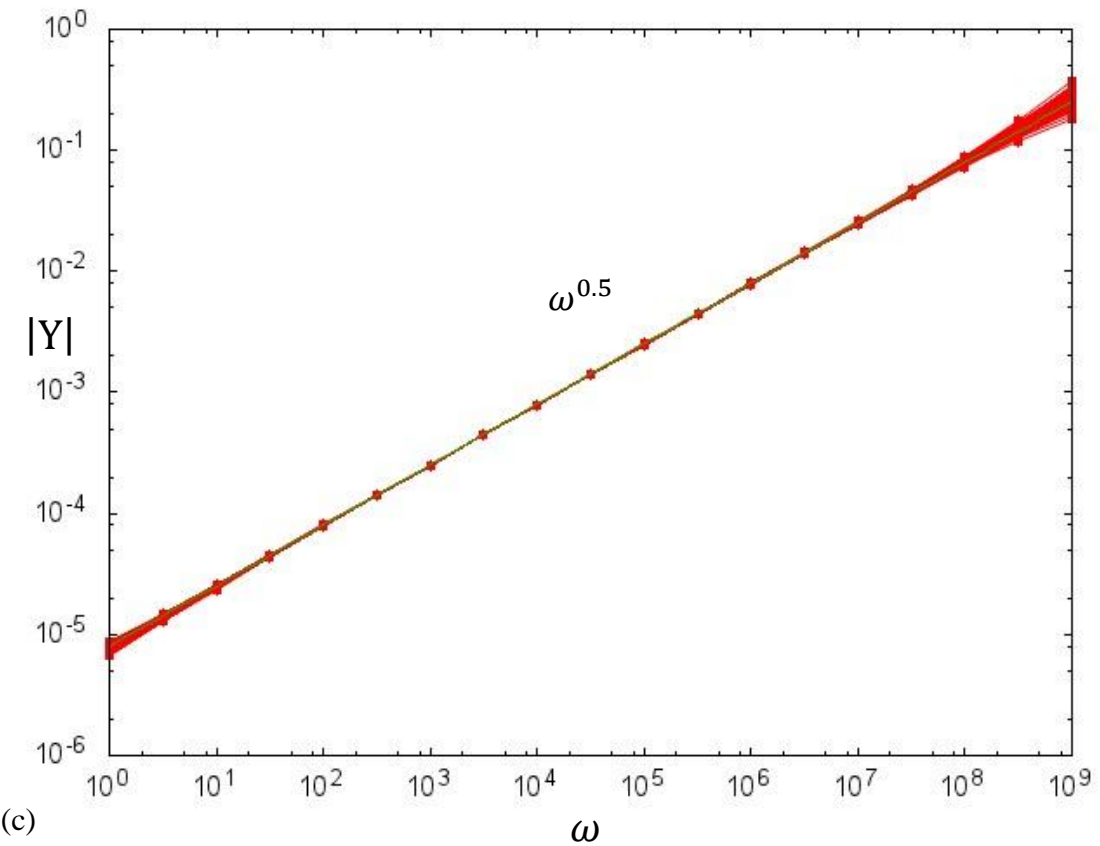

(c)

**Fig. 7.** The overall admittance of network as a function of frequency which contain 2097152 components, with response of 1056 realizations for response of $p = 0.4, 0.6, 0.5$. The green line shows the logarithmic mixing formula, Equation (6), prediction for the networks.